\documentclass[showpacs,pre,onecolumn]{revtex4}
\usepackage{amsmath}
\usepackage{caption2}
\usepackage{graphicx}
\usepackage{subfigure}

\input{tcilatex}
\begin{document}

\title{Two-dimensional solitons in saturable media with a
quasi-one-dimensional lattice potential}
\author{Thawatchai Mayteevarunyoo}
\affiliation{Department of Telecommunication Engineering, Mahanakorn University of
Technology, Bangkok 10530, Thailand}
\author{Boris A. Malomed}
\affiliation{Department of Interdisciplinary Studies, School of Electrical Engineering,
Tel Aviv University, Tel Aviv 69978, Israel}

\begin{abstract}
We study families of solitons in a two-dimensional (2D) model of the light
transmission through a photorefractive medium equipped with a
(quasi-)one-dimensional photonic lattice. The soliton families are bounded
from below by finite minimum values of the peak and total power. Narrow
solitons have a single maximum, while broader ones feature side lobes.
Stability of the solitons is checked by direct simulations. The solitons can
be set in motion across the lattice (actually, made tilted in the spatial
domain), provided that the respective boost parameter does not exceed a
critical value. Collisions between moving solitons are studied too.
Collisions destroy the solitons, unless their velocities are sufficiently
small. In the latter case, the colliding solitons merge into a single stable
pulse.
\end{abstract}

\maketitle

\section{Introduction}

As was first predicted in Refs. \cite{DemetriMoti}, a periodic lattice
potential induced in a photorefractive medium, which is characterized by the
saturable nonlinearity, may be an efficient tool for creation and
stabilization of two-dimensional (2D) spatial solitons of various types. The
prediction was followed by the experimental observation of ordinary solitons 
\cite{Nature}, localized vortices with the topological charge $1$ belonging
to the first (lowest) \cite{vortex} or second \cite{second-band} bandgap in
the respective linear spectrum (stable higher-order vortices and \textit{%
supervortices} in such systems were recently predicted too \cite{Hidetsugu}%
), dipole- and quadrupole-mode solitons \cite{Jianke}, steady patterns in
the form of soliton necklaces \cite{necklace}, and some others. A review of
the field was recently given in Ref. \cite{ExpressReview}.

In the experiment, the periodic potential is induced by the photonic
lattice, which is created as a superposition of counterpropagating laser
beams illuminating the photorefractive crystal in the ordinary polarization,
in which the light propagation is nearly linear; then, the spatial solitons
are built in a probe beam launched through the lattice in the extraordinary
polarization (i.e., polarized along the crystalline $c$ axis), which is
subject to strong nonlinearity induced by the dc bias electric field applied
to the crystal \cite{DemetriMoti,ExpressReview}. Using this technique, the
square photonic lattice can be created by two pairs of counterpropagating
beams illuminating the crystal in directions orthogonal to each other and to
the probe beam that gives rise to the soliton(s).

On the other hand, 2D solitons can also be supported by a low-dimensional,
i.e., quasi-one-dimensional (Q1D) periodic potential, that may be readily
induced by a single pair of counterpropagating beams illuminating the bulk
crystal in the ordinary polarization. Recently, it has been predicted that
the Q1D lattice may efficiently stabilize 2D solitons in the model with the
cubic (rather than saturable) nonlinearity \cite{BBB}. This result directly
applies to Bose-Einstein condensates with attractive interactions between
atoms, trapped in a photonic lattice; moreover, it has also been
demonstrated \cite{BBB,Barcelona} that a quasi-2D lattice can stabilize 3D
solitons against strong collapse that the cubic nonlinearity gives rise to
in the latter case. 2D solitons supported by the Q1D lattice naturally
demonstrate strong anisotropy, which makes them essentially different from
the usual 2D solitons. A significant advantage offered by the use of the Q1D
lattice is the fact that the remaining free direction allows the solitons to
move (in the spatial domain, ``motion" means a tilt of the soliton beam),
which opens a way to study collisions between them, formation of bound
states, etc. \cite{BBB}. The mobility of 2D solitons may also be strongly
anisotropic in some 2D lattices \cite{preprint}.

The objective of the present work is to introduce 2D solitons in the model
of the photorefractive media with the saturable nonlinearity and Q1D
periodic potential (Q1D solitons -- which are, effectively, one-dimensional
objects -- in the model of the 2D photorefractive medium with the Q1D
lattice were introduced in Ref. \cite{Lena}). The soliton solutions are
constructed in Section 2. We demonstrate that the Q1D solitons are
characterized by minimum peak and total intensities necessary for their
existence, and they are stable in the entire existence region. Moving
solitons and collisions between them are studied in Section 3. We find that
collisions destroy the solitons, unless the collision ``velocity" (in fact,
the relative tilt of the two spatial solitons) is small enough; in the
latter case, the colliding solitons merge into a single one, irrespective of
the orientation of the velocity vector relative to the Q1D lattice.

The model outlined above is based on an equation for the spatial evolution
of the probe field (the slowly varying amplitude $U$ of the electromagnetic
wave in the extraordinary polarization), which follows the standard
description of photorefractive media \cite{DemetriMoti}. In normalized
units, the equation takes the form:

\begin{equation}
i\frac{\partial U}{\partial z}+\frac{\partial ^{2}U}{\partial x^{2}}+\frac{%
\partial ^{2}U}{\partial y^{2}}-\frac{U}{1+I_{0}\cos ^{2}\left( \pi
x/d\right) +\left\vert U\right\vert ^{2}}=0,  \label{PhR}
\end{equation}%
where $I_{0}$ and $d$ are the peak intensity and period of the photonic
lattice induced by the superposition of counterpropagating waves launched
along the $x$ axis in the ordinary polarization. The normalized propagation
and transverse coordinates, $z$ and $\left( x,y\right) $, are proportional
to their counterparts measured in physical units, $Z$ and $\left( X,Y\right) 
$, so that $z=Z/\left( k_{0}\Delta n_{0}\right) $ and $\left( x,y\right)
=\left( X,Y\right) /\sqrt{2k_{0}n_{0}\Delta n_{0}}$, where $k_{0}$ is the
propagation constant of the probe wave, and $\Delta n_{0}=n_{0}^{3}r_{33}E/2$
is the change of the refractive index $n_{0}$ (accounted for by the
electro-optic coefficient $r_{33}$) caused by the dc electric field $E$.

\section{Two-dimensional solitons and their stability}

Soliton solutions to Eq. (\ref{PhR}) are searched for as 
\begin{equation}
U=u\left( x,y\right) e^{-i\mu z},  \label{mu}
\end{equation}%
where $-\mu $ is the shift of the propagation constant in the soliton, and
the real function $u$ satisfies the equation%
\begin{equation}
\frac{\partial ^{2}u}{\partial x^{2}}+\frac{\partial ^{2}u}{\partial y^{2}}+%
\left[ \mu -\frac{1}{1+I_{0}\cos ^{2}\left( \pi x/d\right) +u^{2}}\right]
u=0.  \label{NE}
\end{equation}%
Solutions $u(x,y)$ of this equation were found by means of an iteration
procedure in the Fourier space. To this end, following a numerical method
elaborated in Ref. \cite{Jianke}, Eq. (\ref{NE}) was rewritten for the
Fourier transform $\widehat{u}(k_{x},k_{y})$ in the form%
\begin{equation}
\widehat{u}=\frac{1}{k_{x}^{2}+k_{y}^{2}-\mu +1}\left\{ \mathcal{F}\left( 
\frac{I_{0}\cos ^{2}\left( \pi x/d\right) \cdot u}{1+I_{0}\cos ^{2}\left(
\pi x/d\right) +\left\vert u\right\vert ^{2}}\right) +\mathcal{F}\left( 
\frac{\left\vert u\right\vert ^{2}u}{1+I_{0}\cos ^{2}\left( \pi x/d\right)
+\left\vert u\right\vert ^{2}}\right) \right\} ,  \label{Fu}
\end{equation}%
$\mathcal{F}\left( ...\right) $ standing for the 2D Fourier transform. A
direct iteration procedure applied to Eq. (\ref{Fu}) does not converge, in
the general case. Therefore, the equation was modified as follows: defining
integral factors,%
\begin{eqnarray}
\alpha &=&\int \int \left\{ \left( k_{x}^{2}+k_{y}^{2}-\mu +1\right) 
\widehat{u}-\mathcal{F}\left( \frac{I_{l}}{1+I_{0}\cos ^{2}\left( \pi
x/d\right) +\left\vert u\right\vert ^{2}}u\right) \right\} \widehat{u}^{\ast
}dk_{x}dk_{y}  \label{alpha} \\
\beta &=&\int \int \mathcal{F}\left( \frac{\left\vert u\right\vert ^{2}u}{%
1+I_{0}\cos ^{2}\left( \pi x/d\right) +\left\vert u\right\vert ^{2}}\right) 
\widehat{u}^{\ast }dk_{x}dk_{y},  \label{beta}
\end{eqnarray}%
where $\ast $\ stands for the complex conjugation, the following iterative
equation was introduced,%
\begin{eqnarray}
\widehat{u}_{n+1} &=&\frac{1}{k_{x}^{2}+k_{y}^{2}-\mu +1}\left\{ \left( 
\frac{\alpha _{n}}{\beta _{n}}\right) ^{1/2}\mathcal{F}\left( \frac{I_{l}}{%
1+I_{0}\cos ^{2}\left( \pi x/d\right) +\left\vert u_{n}\right\vert ^{2}}%
u_{n}\right) \right.  \notag \\
&&\left. +\left( \frac{\alpha _{n}}{\beta _{n}}\right) ^{3/2}\mathcal{F}%
\left( \frac{\left\vert u_{n}\right\vert ^{2}u_{n}}{1+I_{0}\cos ^{2}\left(
\pi x/d\right) +\left\vert u_{n}\right\vert ^{2}}\right) \right\} ,
\label{iter}
\end{eqnarray}%
where $\alpha _{n}$ and $\beta _{n}$ are the factors (\ref{alpha}) and (\ref%
{beta}) corresponding to the function $\widehat{u}_{n}$. Fixed points of Eq.
(\ref{iter}), corresponding to $\underset{n\longrightarrow \infty }{\lim }%
\left( \alpha _{n}/\beta _{n}\right) =1$, yield solutions to Eq. (\ref{Fu})
as well. The iterative procedure based on Eq. (\ref{iter}) provides for fast
convergence, and produces solutions displayed below.

Typical examples of the 2D solitons are shown in Fig. \ref{fig1}. In
particular, the picture observed in the left column of the figure is typical
to cases when the lattice potential is weak, and/or the soliton's peak
intensity essentially exceeds the lattice's amplitude $I_{0}$: the soliton
is practically isotropic, without a conspicuous effect of the lattice. The
picture in the right column is typical for a relatively strong lattice with
a large period: then, the soliton is almost entirely trapped in one
potential trough, assuming an elliptic shape. In either case, the soliton's
shape features a single lobe, either circular or elliptic one.

The most interesting typical case is represented by the central column in
Fig. \ref{fig1}, which demonstrates a soliton with well-pronounced side
lobes in a moderately strong lattice (in physical units, this case
corresponds to typical values of parameters available in the experiment).
Naturally, the multi-humped structure is observed only along the axis $x$,
while in the free direction, $y$, the soliton always features a simple
single-hump form. Results reported below are given for the same period, $%
d=2\pi $, which gives rise to the central column in Fig. \ref{fig1}. For
other values of $d$ which are neither very small nor very large, the results
are quite similar.

The solitons are characterized by the peak intensity (power), $I_{p}\equiv
\left\vert u(x=0,y=0)\right\vert ^{2}$, and integral intensity,%
\begin{equation}
P\equiv \int_{-\infty }^{+\infty }\int_{-\infty }^{+\infty }\left\vert
u\left( x,y\right) \right\vert ^{2}dxdy.  \label{P}
\end{equation}%
Accordingly, soliton families for given values of $I_{0}$ and $d$ are
represented by the dependences $P(\mu )$ and $I_{p}(\mu )$, see Fig. \ref%
{fig2} [recall $-\mu $ is the soliton propagation-constant shift, defined in
Eq. (\ref{mu})]. An important feature observed in this figure is that the 2D
solitons exist, with a given lattice strength $I_{0}$, only for $P$ and $%
I_{p}$ exceeding certain finite minimum (threshold) values, $P_{\min }$ and $%
I_{p,\min }$; for instance, $I_{p,\min }=0.0411$ for $I_{0}=5$ and $d=2\pi $%
. Both minimum values are shown, as functions of the strength $I_{0}$ of the
photonic lattice, in Fig. \ref{fig3}. It should be noticed that a lower
intensity threshold necessary for the existence of 2D solitons in models
combining lattice potentials with various nonlinearities was found in
earlier works \cite{DemetriMoti,BBBfirst,BBB}.

\begin{figure}[tbp]
\renewcommand{\captionfont}{\small \sffamily} \renewcommand{%
\captionlabelfont}{} \centering%
\subfigure[]{\label{fig:subfig:a}
\includegraphics[width=4in]{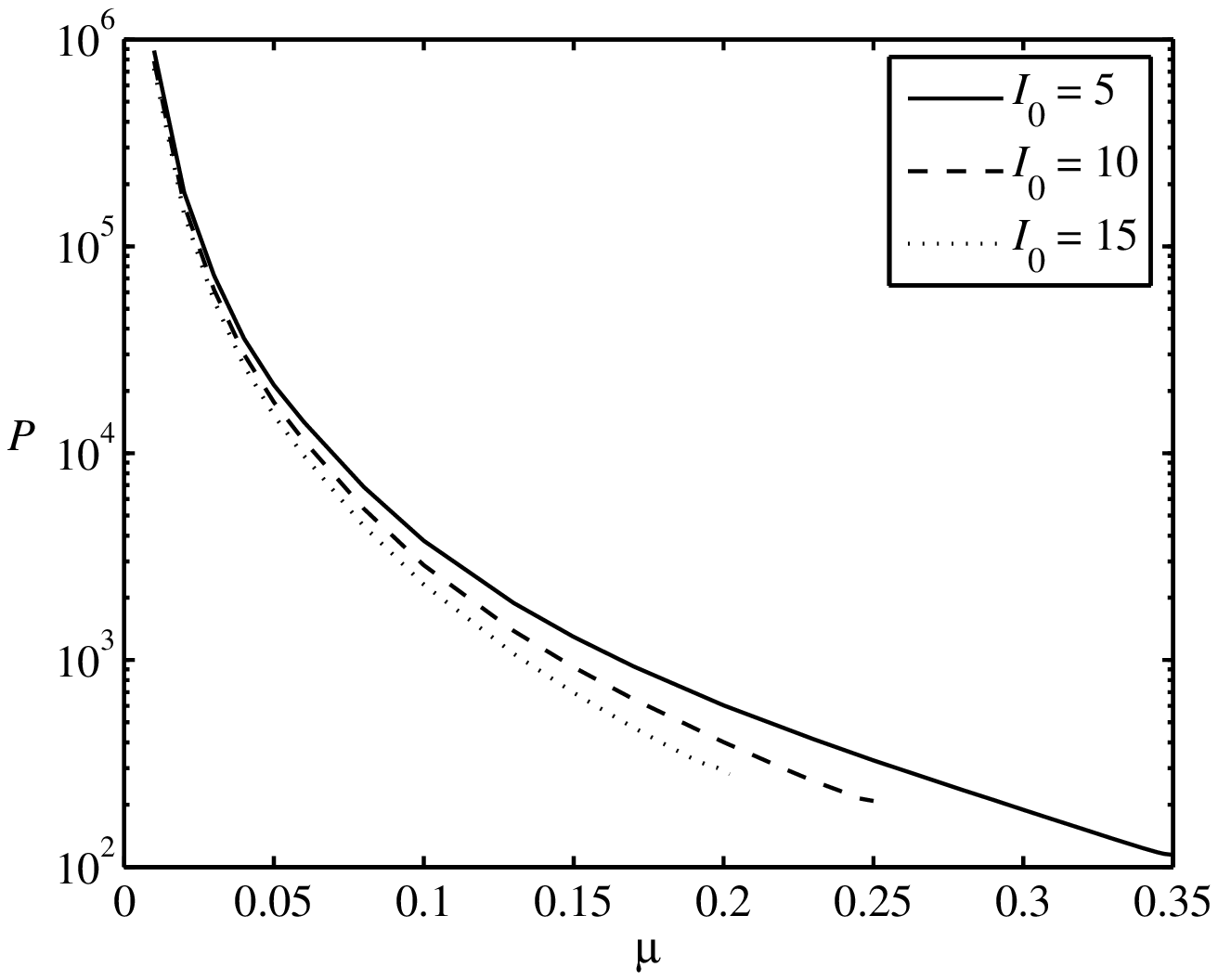}} 
\subfigure[]{\label{fig:subfig:b}
\includegraphics[width=4in]{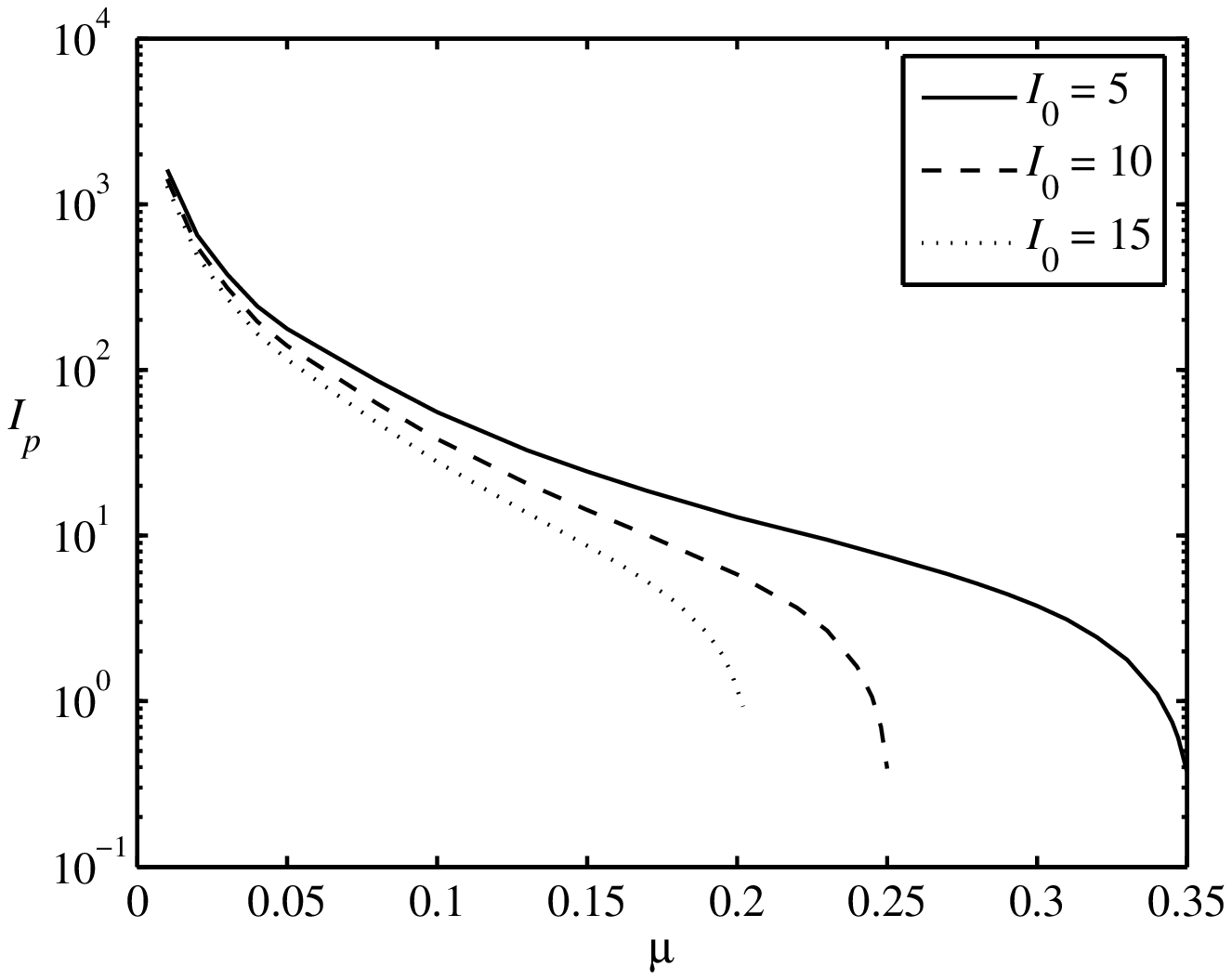}} \renewcommand{\figurename}{Fig.}
\caption{The total intensity (power) (a) and peak intensity (b) of families
of two-dimensional solitons, vs. the absolute value of the
propagation-constant shift, $\protect\mu $, for $d=2\protect\pi $, and
different values of the photonic-lattice strength $I_{0}$. The curves
terminate at points where the solitons cease to exist.}
\label{fig2}
\end{figure}

\begin{figure}[tbp]
\renewcommand{\captionfont}{\small \sffamily} \renewcommand{%
\captionlabelfont}{} \centering\includegraphics[width=4in]{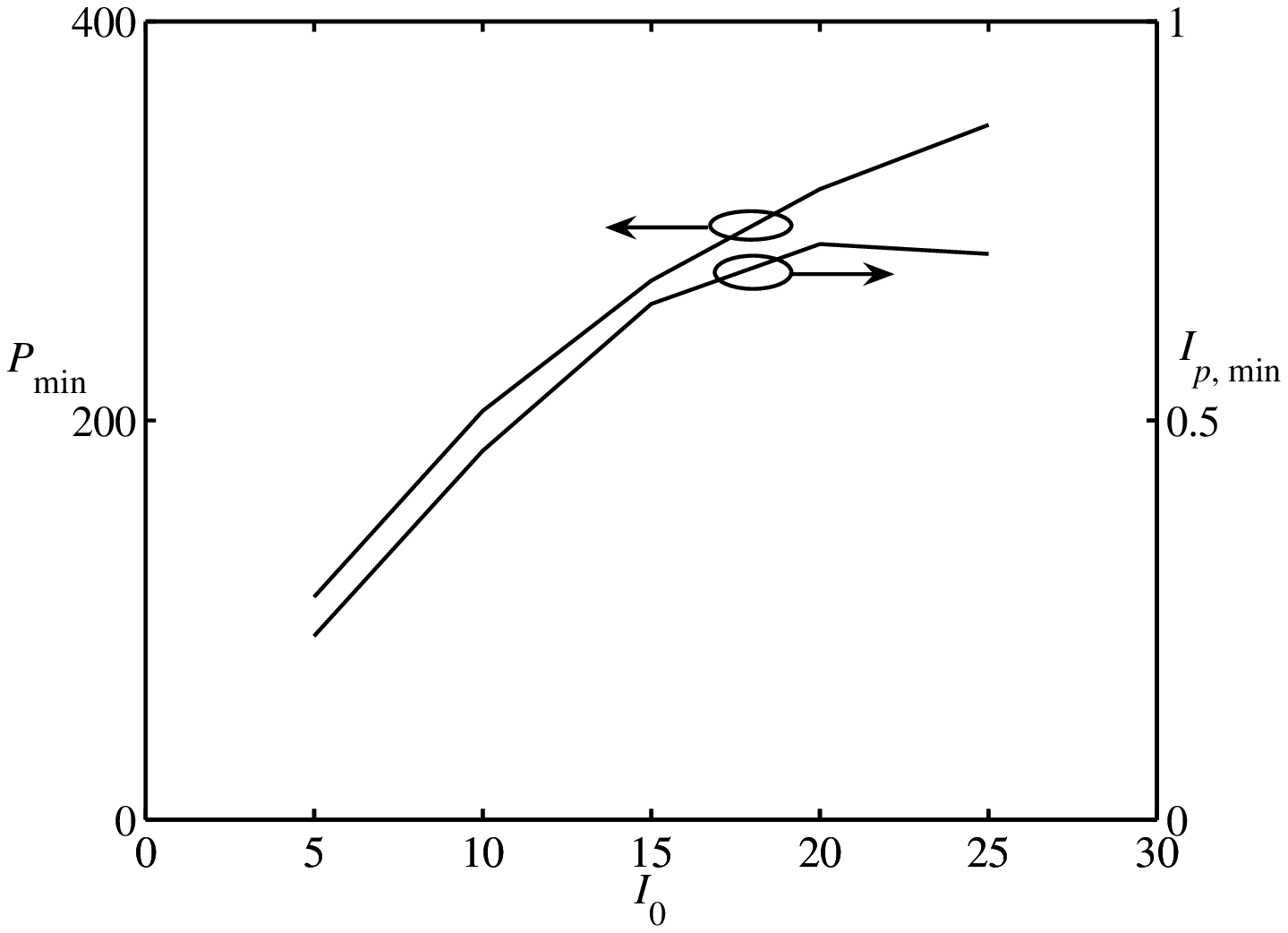} %
\renewcommand{\figurename}{Fig.}
\caption{{}The minimum peak intensity, $I_{p,\min }$, and total power, $%
P_{\min }$, necessary for the existence of the two-dimensional solitons, vs.
the strength, $I_{0}$, of the underlying quasi-one-dimensional lattice.}
\label{fig3}
\end{figure}

It is easy to check that the soliton families shown in Figs. \ref{fig1} and %
\ref{fig2} belong to the semi-infinite bandgap in the spectrum of the
linearized equation (\ref{PhR}). It is known that, on top of the
semi-infinite gap, a sufficiently strong lattice may give rise to solitons
in finite bandgaps of models combining a periodic potential and saturable
nonlinearity (see, e.g., Ref. \cite{Merhasin}). In this work, we do not aim
to look for gap solitons of such types.

Proceeding to the investigation of stability of the 2D solitons, we first of
all note that the negative slope of the dependence $P(\mu )$, obvious in
Fig. \ref{fig2}, suggests possible stability of the soliton families
pursuant to the Vakhitov-Kolokolov (VK)\ criterion \cite{VK}, which may
guarantee the absence of unstable modes of small perturbations corresponding
to real eigenvalues of the instability growth rate. The full stability is
not provided by the VK criterion, and, moreover, the applicability of the
criterion even to perturbations with real eigenvalues was not proven in the
present context. In fact, a counter-example is known: soliton subfamilies
which should be VK-unstable in a 1D model combining a periodic potential of
the Kronig-Penney type and cubic-quintic nonlinearity (another variety of
the saturation) were found to be \emph{completely stable} \cite{Merhasin}.

\begin{figure}[t]
\renewcommand{\captionfont}{\small } \renewcommand{\captionlabelfont}{} %
\centering%
\subfigure[]{\label{fig:subfig:b}
\includegraphics[width=4in]{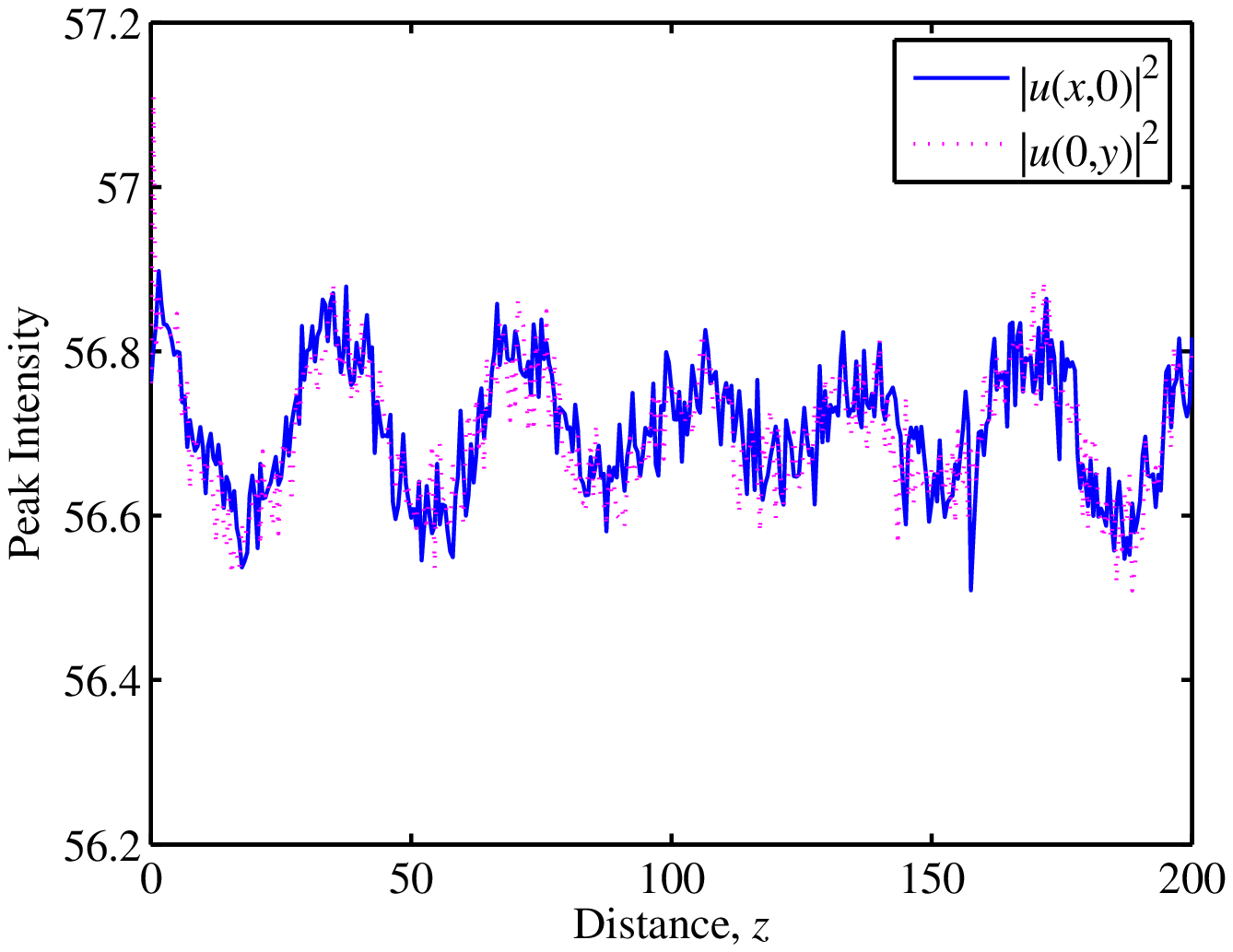}} 
\subfigure[]{\label{fig:subfig:b}
\includegraphics[width=4in]{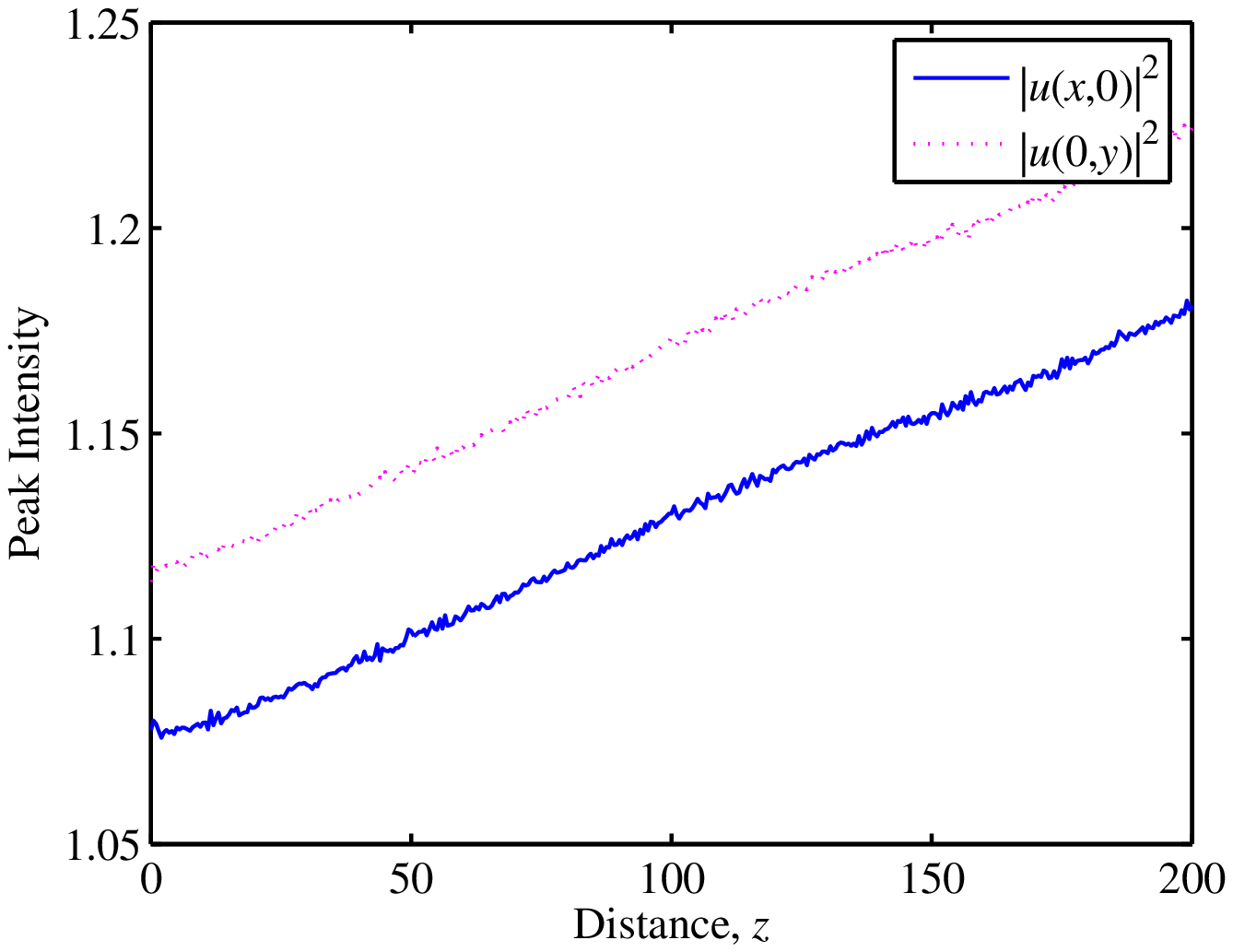}} \renewcommand{\figurename}{Fig.}
\caption{{}(a) Stable evolution of a high-intensity soliton, with $I_{0}=5$
and $I_{p}=56.76$, under random perturbations, is illustrated by the $z$%
-dependence of the soliton's peak intensity. (b) The same for a
low-intensity soliton, with $I_{p}=1.1$.}
\label{fig4}
\end{figure}

We have tested stability of the 2D solitons in direct simulations of Eq. (%
\ref{PhR}), making use of the fast Fourier transform in directions $x$ and $%
y $ and Runge-Kutta method to advance in $z$. In each simulation, some
amount of random noise was added to the soliton as a perturbation. It has
been concluded that \emph{all the solitons} are stable, as (formally)
predicted by the VK criterion. An example illustrating the stability of a
large-amplitude 2D soliton with $I_{0}=5$ and $I_{p}=56.76$ is displayed in
Fig. \ref{fig4}(a). Broad solitons, with the peak intensity close to the
threshold value $I_{p,\min }$, are stable too, although their peak intensity
may slowly grow with $z$, as shown in Fig. \ref{fig4}(b) for the same
lattice strength, $I_{0}=5$, as in Fig. \ref{fig4}(a), but $I_{p}=1.1$. A
plausible explanation to the latter effect is that the combination of the
lattice potential and self-focusing nonlinearity (for relatively small $%
I_{p} $, the saturation does not suppress the self-focusing) leads to
sucking the perturbation wave field into the spot where the soliton's
maximum is located, and the build-up of the additional wave field around the
soliton's peak may be conspicuous, against the backdrop of the relatively
small value of $I_{p}$.

\subsection{Moving two-dimensional solitons and their collisions}

Solutions $\tilde{U}(x,y,z)$ for solitons ``moving" (actually, tilted) along
the free direction $y$ can be generated from the ``quiescent" solitons
reported in the previous section, $U(x,y,z)$, by means of the Galilean
transformation,%
\begin{equation}
\tilde{U}(x,y,z)=U(x,y-2Q_{y}z,z)\exp \left( iQ_{y}x-iQ_{y}^{2}z\right) ,
\label{Galileo}
\end{equation}%
where $Q_{y}$ is an arbitrary \textit{boost parameter}, that determines the
soliton's ``velocity" (tilt) $2Q_{y}$. On the other hand, generation of
solitons tilted along the $x$ axis is a nontrivial problem. In the 1D
version of Eq. (\ref{PhR}), without the term $\partial ^{2}U/\partial y^{2}$%
, a family of tilted solitons was introduced in Ref. \cite{Lena}.

We looked for solitons ``moving" along the $x$ axis by simulating the
evolution of initial states of the form%
\begin{equation}
U_{0}(x,y)=U(x,y)\exp \left( iQ_{x}x\right) ,  \label{U0}
\end{equation}%
cf. Eq. (\ref{Galileo}), where $U(x,y)$ corresponds to a zero-velocity
stationary soliton solution. The boost parameter in Eq. (\ref{U0}), $Q_{x}$,
was gradually increased from a run to a run, until reaching a critical
(maximum) value $Q_{\max }$, at which the initial configuration (\ref{U0})
does not generate any soliton, but rather gets destroyed into
small-amplitude waves. Results of the simulations are summarized in Fig. \ref%
{fig5}, in the form of plots showing $Q_{\max }$ as a function of the
soliton's peak intensity $I_{p}$, for different values of the lattice
strength $I_{0}$. Measurement of the established value of the average
``velocity" (tilt) of the solitons generated by the initial condition (\ref%
{U0}) with $Q_{x}<Q_{\max }$ produces values which are quite close to ones ($%
2Q_{x}$) corresponding to the Galilean transform (\ref{Galileo}) in the free
space. Finally, applying the Galilean transformation (\ref{Galileo}) to the
soliton already moving along the $x$ direction, one can generate a pulse
moving in any direction (in particular, along the diagonal, $x=y$, see
below).

Once the stability limits for the moving solitons are available, the next
step is to consider collisions between them. The outcome of the collision
may depend on the magnitude and direction of the velocities, and the aiming
mismatch (its zero value corresponds to the head-on collision).

Numerous simulations demonstrate that the collision completely destroys both
solitons, unless their velocities are sufficiently small. An example of the
destructive collision is displayed in Fig. \ref{fig6}. In particular, in the
case of $I_{0}=5$, the solitons must not be set in motion by the boost with $%
\left\vert Q_{x}\right\vert $ or $\left\vert Q_{y}\right\vert $ in excess of 
$0.05,$ otherwise the collision will destroy them. This critical value is
the same for the motion in the $x$ and $y$ directions, up to the accuracy of
the simulations. Moreover, if the solitons are boosted in the diagonal
direction and then collide, the same limit, $\left\vert Q\right\vert =0.05$,
was found for the absolute value of the corresponding vectorial boost, $%
\left\vert Q\right\vert =\sqrt{Q_{x}^{2}+Q_{y}^{2}}$, i.e., the critical
boost is virtually isotropic.

Collisions between solitons boosted by $|Q|$ with values smaller than the
critical one result in their \emph{merger} into a single pulse. An example
of the merger resulting from the collision with a finite aiming mismatch is
displayed in Fig. \ref{fig7}. Further, the mergers caused by head-on
collisions in the diagonal or vertical ($y$) direction are presented in Fig. %
\ref{fig8}. As seen from the figures, the pulse generated by the merger
performs intrinsic vibrations, but remains stable.

The above examples displayed collisions of single-lobe solitons. Collisions
between their broader counterparts, which feature side lobes in the $x$%
-direction, are quite similar. Figure \ref{fig9} presents an example of the
head-on collision between sufficiently slowly moving solitons of the latter
type. The collision again leads to the merger of the pulses into a single
one, which then performs conspicuous intrinsic vibrations, but remains a
stable object. Additional simulations show that the soliton produced by the
merger of two multi-lobe ones can also easily move across the lattice, if
given a push.

Thus we conclude that collisions between the 2D solitons are always \emph{%
strongly inelastic} (both the destruction and merger of the colliding
solitons are inelastic outcomes), attesting to the fact that the present
model is far from any integrable limit, where collisions between solitons
would be elastic. For comparison, we note that the collisions may be less
inelastic in the 2D model with the Q1D periodic potential if the
nonlinearity is cubic \cite{BBB}. In that model, elastic collisions are
possible (passage of the solitons in the case of a finite aiming mismatch,
or their mutual bounce after the head-on collision, if the phase shift
between them is $\pi $). An additional inelastic outcome of the head-on
collisions between in-phase 2D solitons, which was observed in the model
with the cubic nonlinearity, but cannot occur if the nonlinearity is
saturable, is collapse of the single pulse formed after the merger of two
solitons (i.e., formation of a singularity after a finite propagation
distance).

\section{Conclusion}

In this paper, we have proposed a model of the two-dimensional (2D) medium
with the saturable nonlinearity and quasi-1D lattice potential, that can be
realized in the spatial domain in a bulk photorefractive crystal. The
subject of the analysis were 2D solitons (ones belonging to the
semi-infinite bandgap in the linear spectrum). It was demonstrated that they
form families bounded from below by finite minimum values of the peak and
total intensities. Narrow pulses feature a single maximum, while broad
solitons contain side lobes. Direct simulations confirm that the solitons
are stable. They can be set in motion (actually, tilted in the spatial
domain) in an obvious way along the quasi-1D lattice, and also across the
lattice, provided that, in the latter case, the boost parameter does not
exceed a critical value, beyond which the soliton is destroyed.

Collisions between stable moving solitons were studied in detail, with the
conclusion that the collisions destroy the solitons, unless their velocities
are sufficiently small. In the latter case, the colliding solitons merge
into a single stable pulse, that performs intrinsic vibrations. The
predictions reported in this paper can be readily implemented experimentally
in photorefractive crystals.

\begin{figure}[tbp]
\renewcommand{\captionfont}{\small \sffamily} \renewcommand{%
\captionlabelfont}{} \centering\includegraphics[width=4in]{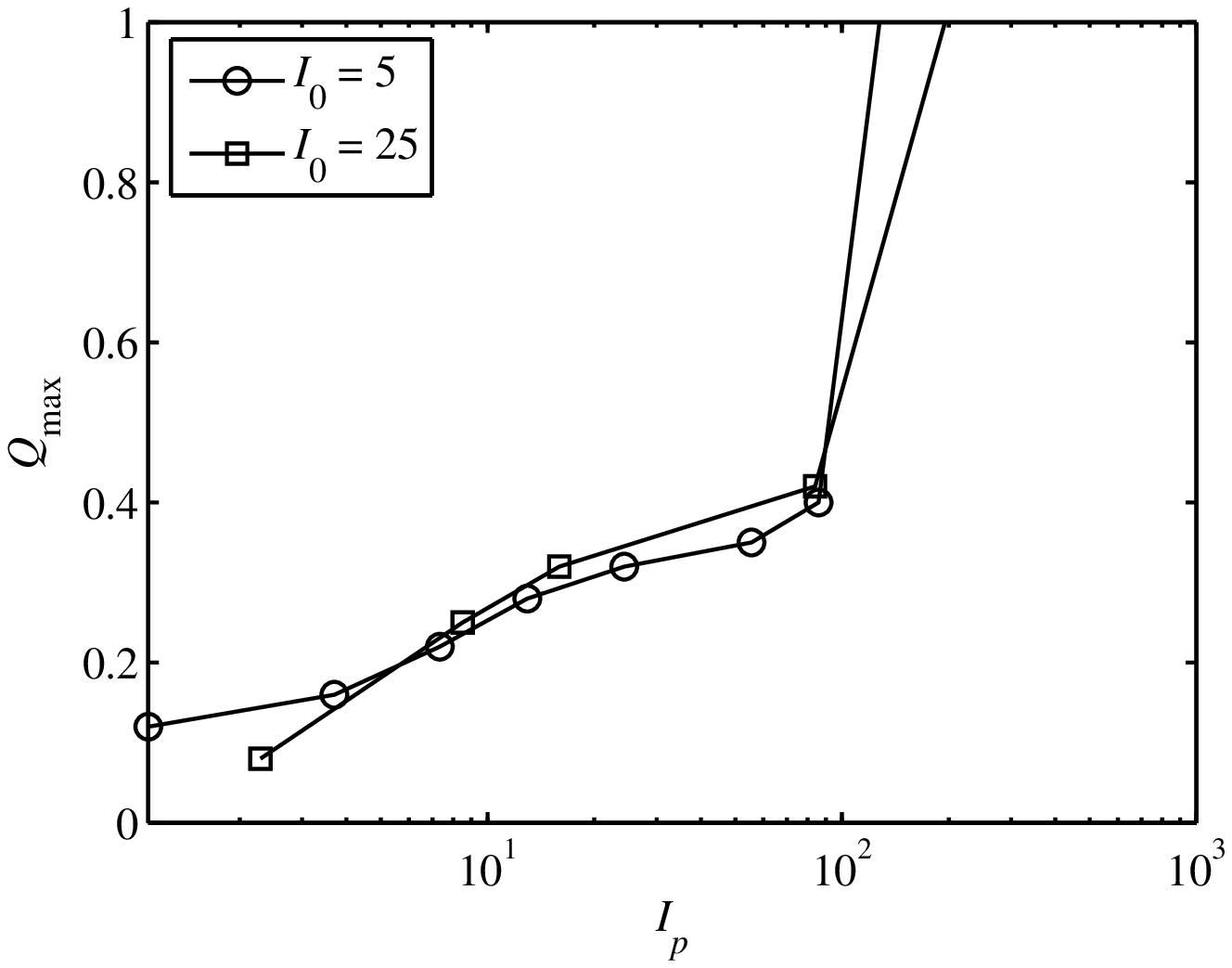} %
\renewcommand{\figurename}{Fig.}
\caption{The maximum value of the boost parameter, $Q_{x}$, which generates
solitons moving (actually, tilted in the spatial domain) along the $x$ axis,
vs. the soliton's peak intensity, $I_{p}$, for different value of the
strength $I_{0}$ of the underlying lattice. For $Q_{x}>Q_{\max }$, the
application of the boost destroys the soliton.}
\label{fig5}
\end{figure}
\ 

\begin{figure}[tbp]
\renewcommand{\captionfont}{\small \sffamily} \renewcommand{%
\captionlabelfont}{} \centering\includegraphics[width=4in]{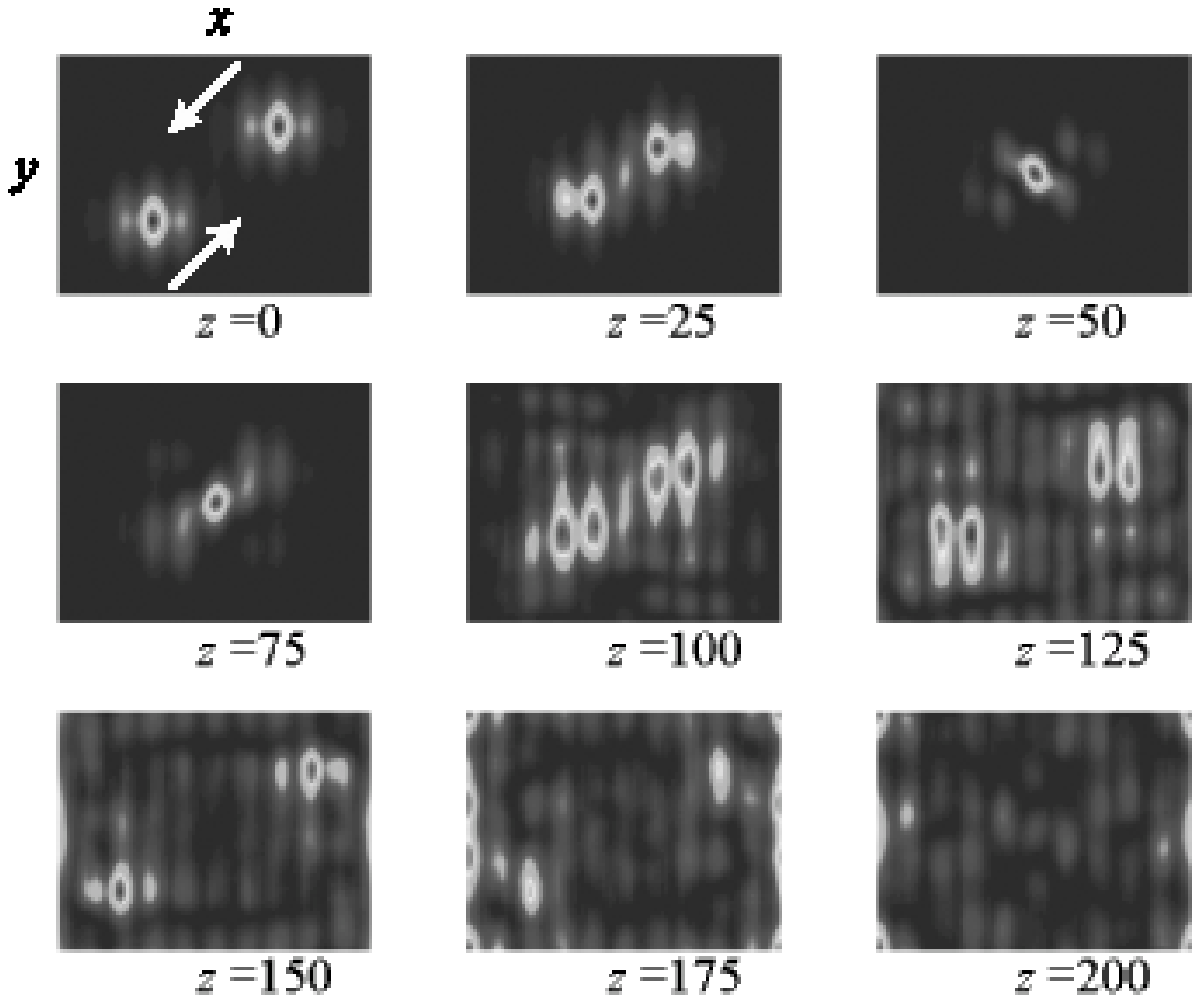} %
\renewcommand{\figurename}{Fig.}
\caption{A typical example of destructive collisions between moving
solitons, for $I_{0}=5$ and $I_{p}=1.78$. The velocity vectors of the
colliding solitons (shown by arrows in the first panel, in this figure and
below) have components $Q_{x,y}=\pm 0.12$.}
\label{fig6}
\end{figure}

\begin{figure}[tbp]
\renewcommand{\captionfont}{\small \sffamily} \renewcommand{%
\captionlabelfont}{} \centering
\includegraphics[width=4in]{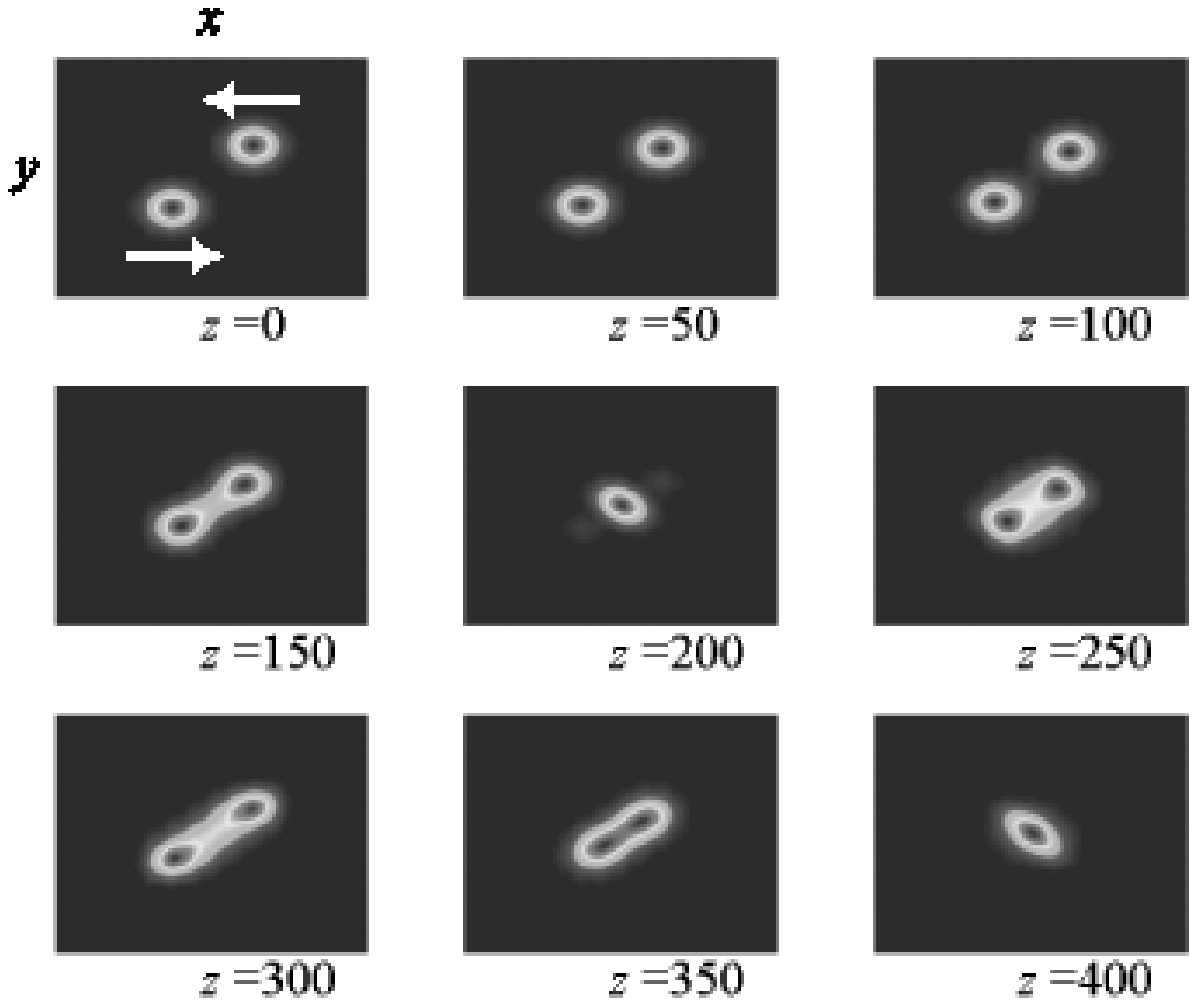} 
\caption{Merger of colliding solitons which were set in motion by the boost $%
Q_{x}=\pm 0.01$, with a finite aiming mismatch in the $y$ direction. In this
case, $I_{0}=25$ and $I_{p}=83.97.$}
\label{fig7}
\end{figure}

\begin{figure}[tbp]
\renewcommand{\captionfont}{\small \sffamily} \renewcommand{%
\captionlabelfont}{} \centering%
\subfigure[]{\label{fig:subfig:a}
\includegraphics[width=4in]{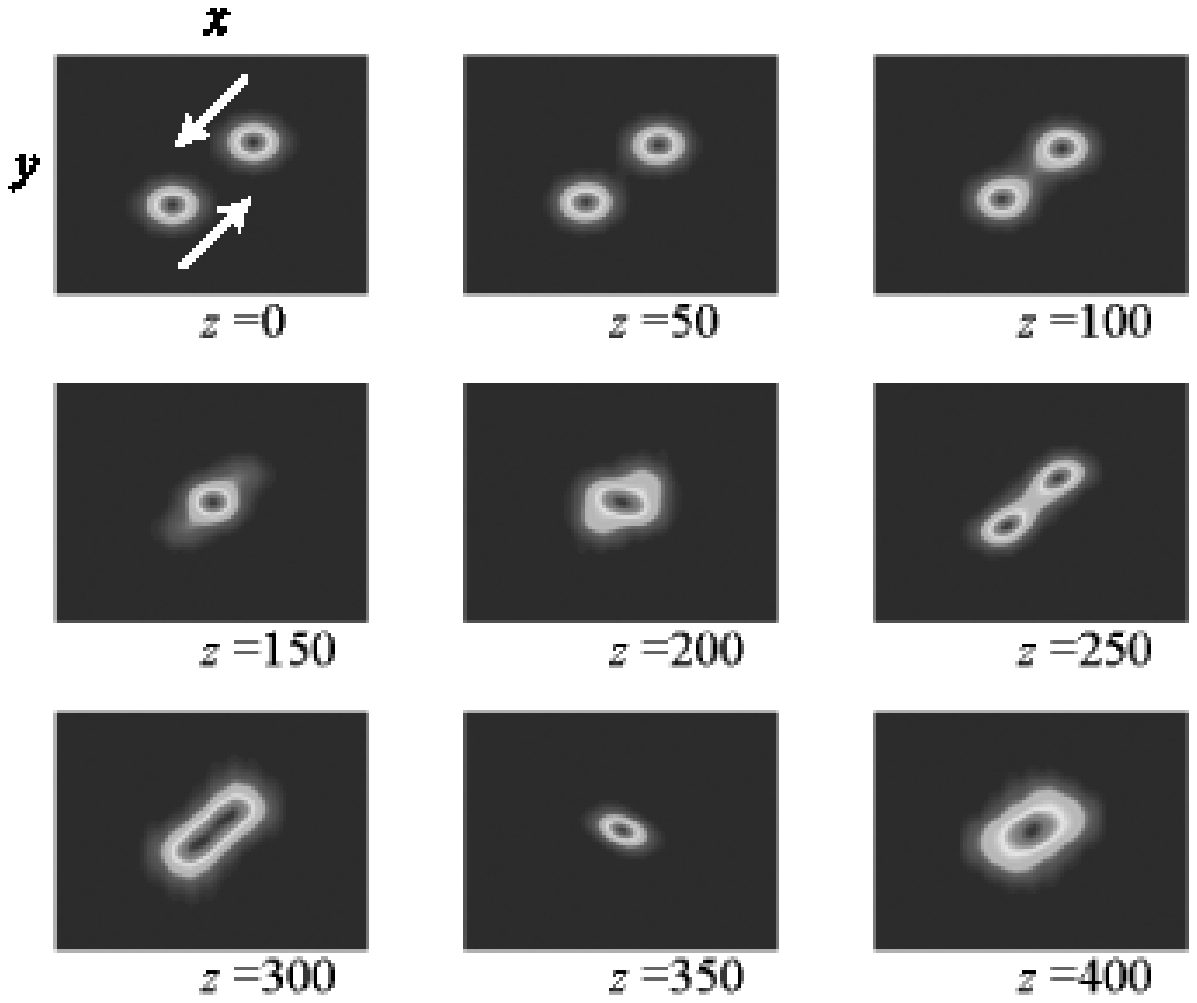}} 
\subfigure[]{\label{fig:subfig:b}
\includegraphics[width=4in]{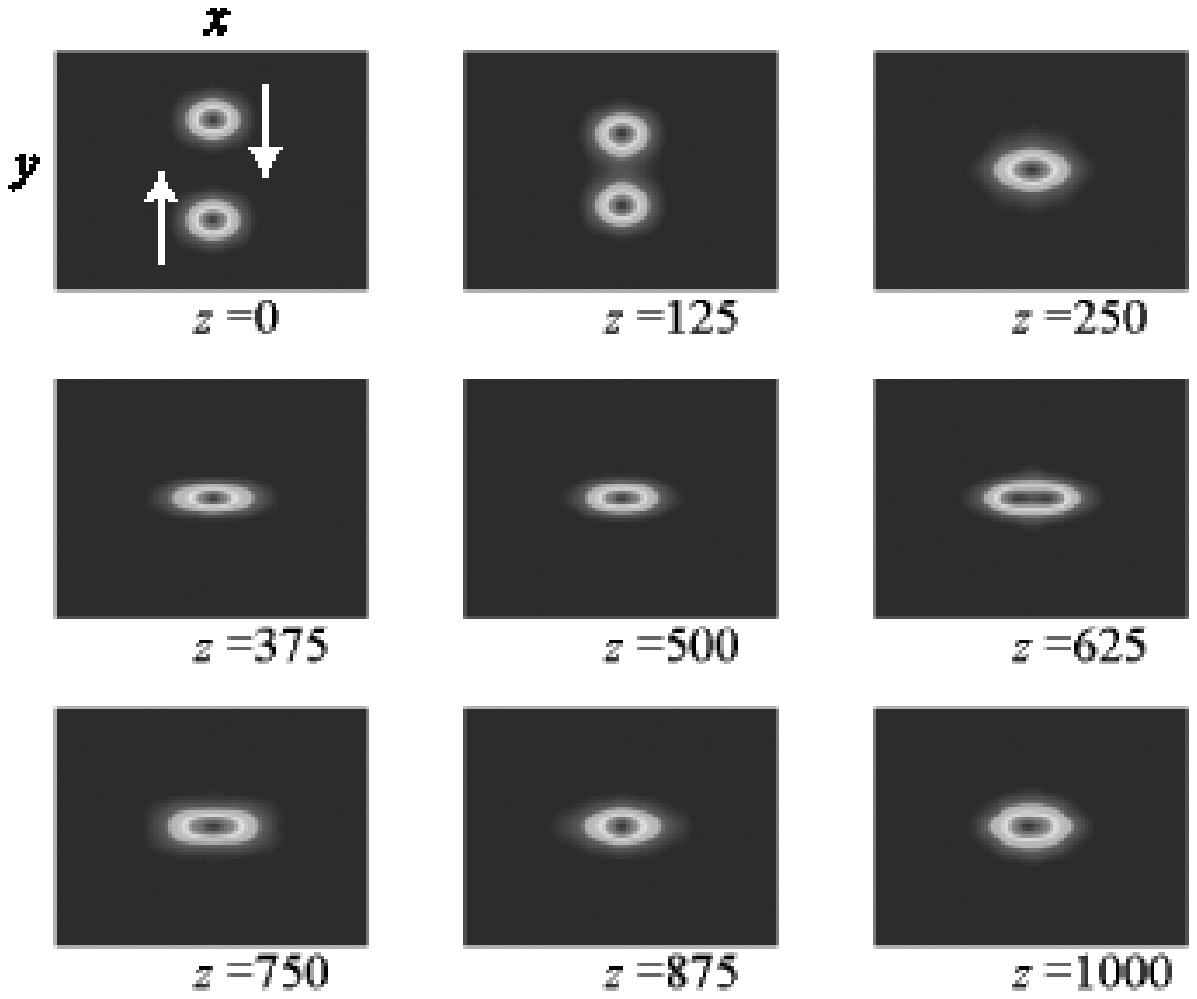}} 
\caption{{}Merger as a result of head-on collisions between slowly moving
solitons that were set in motion (a) along the diagonal direction, by the
application of the boost with $Q_{x}=Q_{y}=\pm 0.01$, and (b) in the
vertical direction, by the boost $Q_{y}=\pm 0.01$. In case (a), $I_{0}=25$
and $I_{p}=83.97$; in case (b), $I_{0}=15$ and $I_{p}=27.92$.}
\label{fig8}
\end{figure}

\begin{figure}[tbp]
\renewcommand{\captionfont}{\small \sffamily} \renewcommand{%
\captionlabelfont}{} \centering\includegraphics[width=4in]{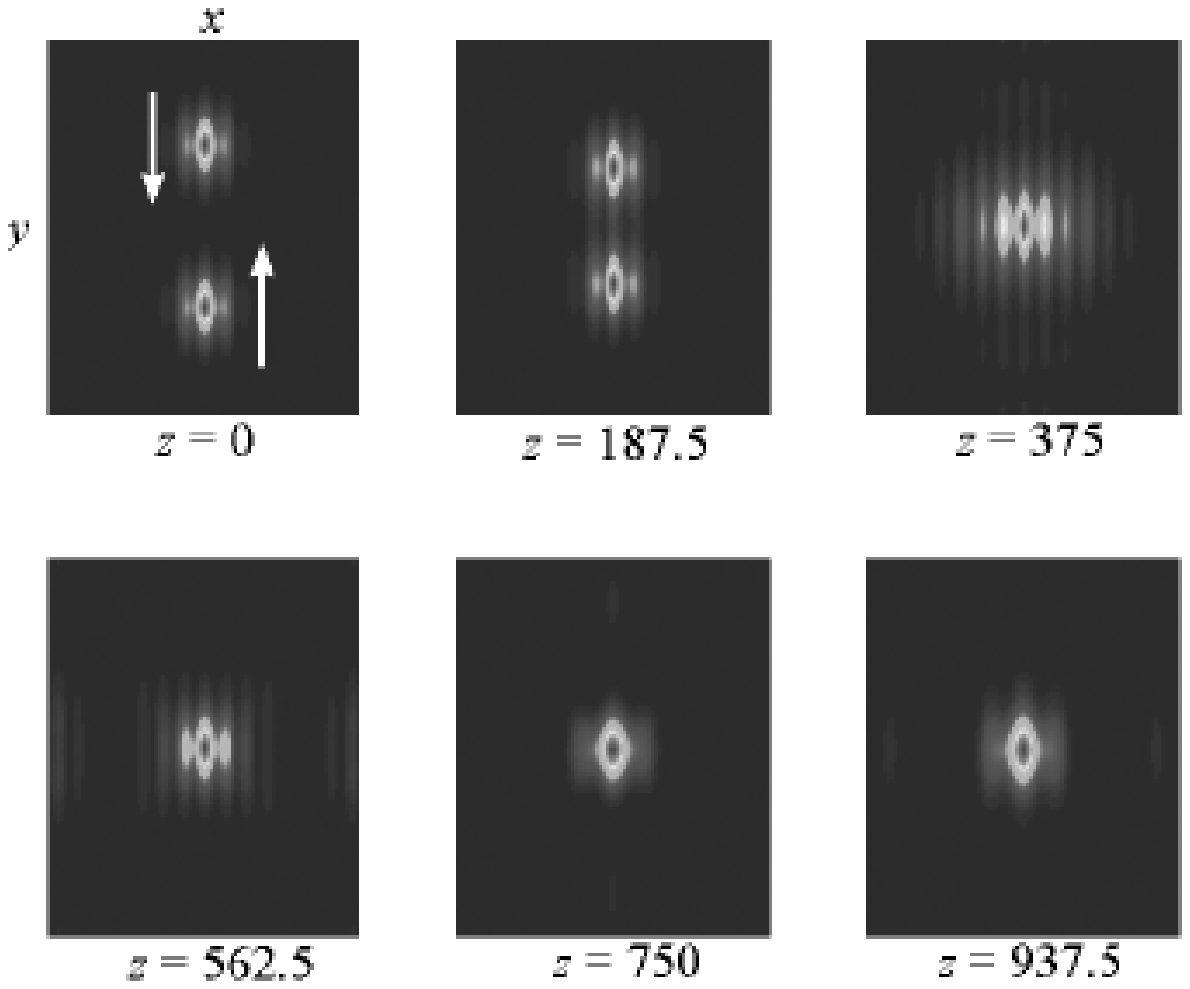} 
\caption{{}Merger of broad (multi-lobed) solitons caused by the head-on
collision. The solitons were set in motion by the vertical boost with $%
Q_{y}=\pm 0.01$. In this case, $I_{0}=5$ and $I_{p}=1.786$.}
\label{fig9}
\end{figure}


\begin{thebibliography}{99}
\bibitem{DemetriMoti} N. K. Efremidis, S. Sears, D. N. Christodoulides, J.
W. Fleischer, and M. Segev, Phys. Rev. E \textbf{66}, 046602 (2002); N. K.
Efremidis, J. Hudock, D. N. Christodoulides, J. W. Fleischer, O. Cohen, and
M. Segev, Phys. Rev. Lett. \textbf{91}, 213906 (2003).

\bibitem{Nature} J.~W. Fleischer, M. Segev, N.~K. Efremidis, and D.~N.
Christodoulides, Nature \textbf{422}, 147 (2003); J. W. Fleischer, G.
Bartal, O. Cohen, O. Manela, M. Segev, J. Hudock, and D. N. Christodoulides,
Phys. Rev. Lett. \textbf{92}, 123904 (2004).

\bibitem{vortex} D N. Neshev, J. Alexander, E. A. Ostrovskaya, Y. S.
Kivshar, H. Martin, I. Makasyuk, and Z. Chen, Phys. Rev. Lett. \textbf{92},
123903 (2004); J. W. Fleischer, G. Bartal, O. Cohen, O. Manela, M. Segev, J.
Hudock, and D. N. Christodoulides, Phys. Rev. Lett. \textbf{92}, 123904
(2004).

\bibitem{second-band} O. Manela, O. Cohen, G. Bartal, J. W. Fleischer, and
M. Segev, Opt. Lett. \textbf{29}, 2049 (2004).

\bibitem{Hidetsugu} H. Sakaguchi and B. A. Malomed, \textit{Higher-order
vortex solitons, multipoles, and supervortices on a square optical lattice},
Europhys. Lett., in press.

\bibitem{Jianke} J. Yang, I. Makasyuk, A. Bezryadina, and Z. Chen, Stud.
Appl. Math. \textbf{113}, 389 (2004).

\bibitem{necklace} J. Yang, I. Makasyuk, P. G. Kevrekidis, H. Martin, B. A.
Malomed, D. J. Frantzeskakis, and Z. Chen, Phys. Rev. Lett. \textbf{94},
113902 (2005).

\bibitem{ExpressReview} J. W. Fleischer, G. Bartal, O. Cohen, T. Schwartz,
O. Manela, B. Freedman, M. Segev, H. Buljan, N. K. Efremidis, Opt. Exp. 
\textbf{13,} 1780 (2005).

\bibitem{BBB} B. B. Baizakov, B. A. Malomed and M. Salerno, Phys. Rev. A 
\textbf{70}, 053613 (2004); in: \textit{Nonlinear Waves: Classical and
Quantum Aspects}, ed. by F. Kh. Abdullaev and V. V. Konotop, p. 61 (Kluwer
Academic Publishers: Dordrecht, 2004; also available at
http://rsphy2.anu.edu.au/\symbol{126}asd124/Baizakov\_2004\_61%
\_NonlinearWaves.pdf)

\bibitem{Barcelona} D. Mihalache, D. Mazilu, F. Lederer, Y. V. Kartashov,
L.-C. Crasovan, and L. Torner, Phys. Rev. E 70, 055603(R) (2004).\newline

\bibitem{preprint} R. Fischer, D. Traeger, D. N. Neshev, A. A. Sukhorukov,
W. Krolikowski, C. Denz, and Y. S. Kivshar, e-print physics/0509255 (2005).

\bibitem{Lena} B. A. Malomed, T. Mayteevarunyoo, E. A. Ostrovskaya, and Y.
S. Kivshar, Phys. Rev. E \textbf{71}, 056616 (2005).

\bibitem{BBBfirst} B. B. Baizakov, B. A. Malomed, and M. Salerno, Europhys.
Lett. \textbf{63}, 642 (2003).

\bibitem{Merhasin} I. M. Merhasin, B. V. Gisin, R. Driben, and B. A.
Malomed, Phys. Rev. E \textbf{71}, 016613 (2005).

\bibitem{VK} M. G. Vakhitov and A. A. Kolokolov, Izv. Vuz. Radiofiz. \textbf{%
16,} 1020 (1973) [in Russian; English translation: Sov. J. Radiophys.
Quantum Electr. \textbf{16}, 783 (1973)]; see also L. Berg{\'{e}}, Phys.
Rep. \textbf{303}, 260 (1998).

\begin{figure}[tbp]
\renewcommand{\captionfont}{\small \sffamily} \renewcommand{%
\captionlabelfont}{} \centering\includegraphics[width=4in]{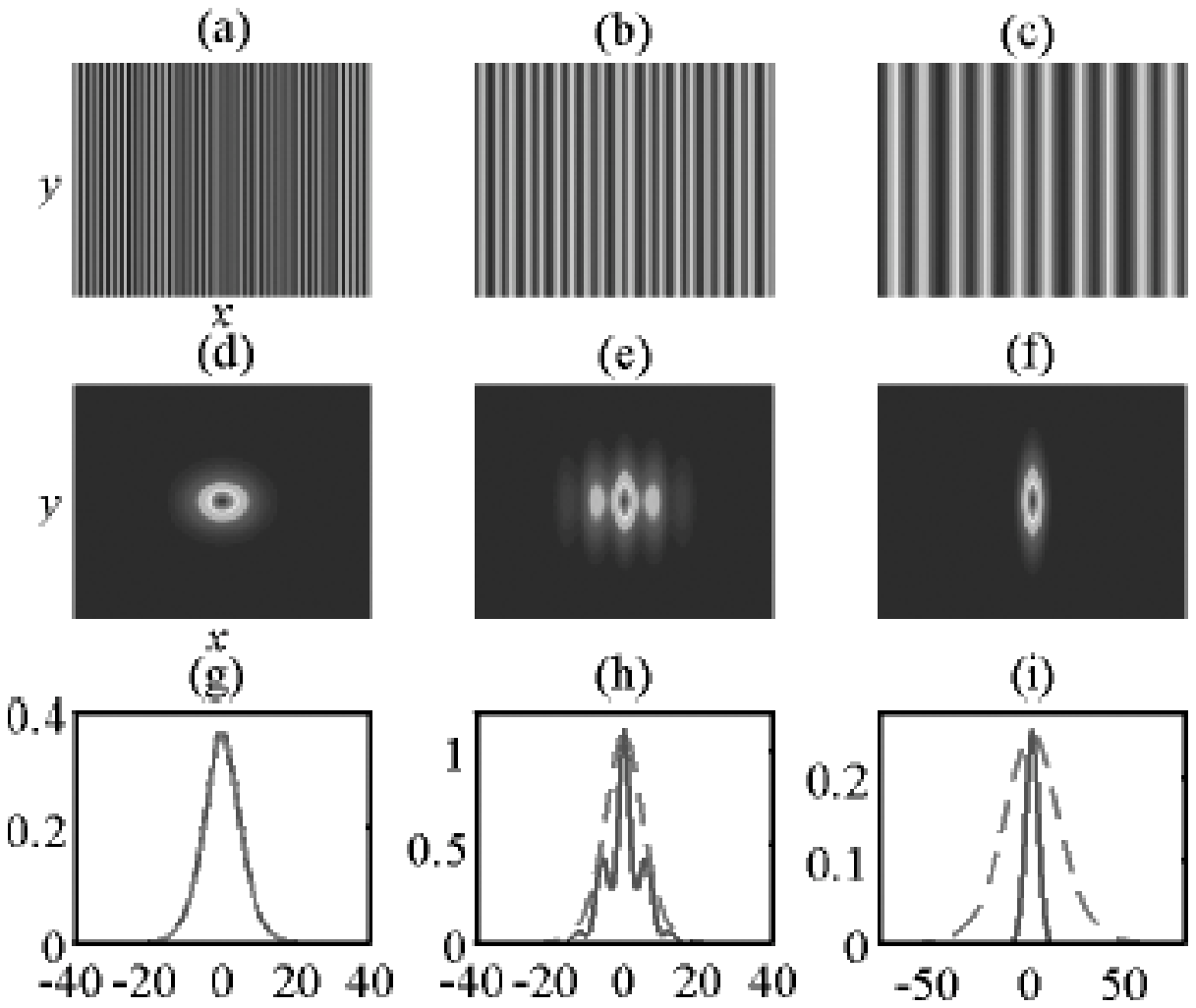} %
\renewcommand{\figurename}{Fig.}
\caption{Typical examples of two-dimensional solitons found from Eq. (%
\protect\ref{NE}) with a moderately strong quasi-one-dimensional photonic
lattice, corresponding to $I_{0}=5$. Panels (a), (b), and (c) show the
intensity distribution in the lattice, $I_{\mathrm{phl}}(x)=I_{0}\cos
^{2}\left( \protect\pi x/d\right) $, for $d=\protect\pi /5$, $2\protect\pi $%
, and $10\protect\pi $, respectively. The two-dimensional intensity field in
the respective solitons is shown in panels (d), (e), and (f), and the
corresponding intensity profiles in two cross sections of the solitons,
along $y=0$ and $x=0$, are displayed in panels (g), (h), and (i).}
\label{fig1}
\end{figure}
\end{thebibliography}
\end{document}